\documentclass[conference]{IEEEtran}

\usepackage{multirow}
\usepackage{amsfonts}
\usepackage{amsmath}
\usepackage{color}
\usepackage{bm}
\usepackage{amssymb}
\usepackage{graphicx}
\usepackage{algorithmicx}
\usepackage{algpseudocode}
\usepackage{algorithm}
\usepackage{fancyhdr}
\usepackage{epstopdf}
\usepackage{paralist}
\usepackage{url}
\usepackage{subfigure}
\usepackage{changes}
\usepackage{cite}

\colorlet{Changes@Color}{green}

\IEEEoverridecommandlockouts

\makeindex
\begin{document}

\title{Adaptive PMU-based Distribution System State Estimation exploiting the Cloud-based IoT paradigm}

\author{\IEEEauthorblockN{Paolo Attilio Pegoraro, Alessio Meloni, Luigi Atzori, Paolo Castello, Sara Sulis\thanks{\copyright~2016 IEEE. The IEEE copyright notice applies.}}
\IEEEauthorblockA{DIEE, University of Cagliari, Italy\\
 \textit{\{paolo.pegoraro,alessio.meloni,l.atzori,paolo.castello,sara.sulis\}@diee.unica.it}\\
}

%\thanks{This work has been supported by Regione Autonoma della Sardegna, L.R. 7/2007: Promozione 
%della ricerca scientifica e dell'innovazione tecnologica in Sardegna, annualit\`{a} 2012, 
%CRP-60511".}

}

\maketitle

\begin{abstract}
This paper presents an adaptive Distribution System State Estimation (DSSE) which relies on a Cloud-based IoT paradigm.
The methodology is adaptive in terms of the rate of execution of the estimation process which varies depending on the indications of the distributed measurement system.
The system is composed, in particular, of Phasor Measurement Units (PMUs).
PMUs are virtualized with respect to the physical devices and the corresponding virtualizing modules run in the 
communication network edge (i.e. closer to the physical objects).
PMUs are set at a higher measurement rate, while the estimation process works at a given slower 
rate, for example once per 
second, in normal operative conditions.
A local decision algorithm implemented in the virtualized module, monitors the measured quantities in order to detect and address possible unexpected dynamics.
In particular, different metrics can be applied: the variations and the trend of variation of the rms voltage values, 
but also the Rate Of Change Of Frequency (ROCOF) of the monitored signals can be used to trigger rate variation in the DSSE.
In case dynamics are detected, the measurement data is sent to the DSSE at higher rates and the estimation process runs consequently on a 
finer time scale.
In the considered system only application level entities are located in the Cloud, thus allowing to obtain a bandwidth-efficient and smart 
data transmission. The results obtained on a 13-bus systems prove the goodness of the proposed methodologies.
\end{abstract}


\section{Introduction}
% no \IEEEPARstart
%The power systems are mainly divided into three parts: the generation system, the transmission network, and the Distribution System (DS).
%They were designed to transport energy from a limited number of traditional power plants towards a great amount of consumers, with a unidirectional energy flow (from generators to loads) and a unidirectional information flow (from users to operators). 
Typical Distribution Systems (DSs) are composed of a very high number of nodes, balanced and unbalanced lines and few measurement devices for the monitoring of the quantities of interest.
The traditional topologies are open-loop (rare reconfiguration possibilities), with substantially radial or only weakly meshed structures.
Nevertheless, DSs are evolving towards more complex and dynamic architectures. 
In particular, the growing presence of Distributed Energy Resources (DERs), including small-scale 
generators, often based on Renewable Energy Sources (RESs), electric vehicles (EVs), and new 
high-efficiency residential and commercial appliances is a crucial event in grids evolution 
\cite{Heydt_NextGenDistrSyst}.

The increasing quantities of DERs located throughout the DSs are drivers and challenges for the distribution business. 
%DERs bring both economic and environmental benefits. 
However, for their own nature, most DERs work in a substantially unpredictable manner and may cause bi-directional power flows and voltage transient problems. 
Grid stability is based on inertia of large rotating machinery, but most \textit{green} energy sources lack short-term inertia. 
They can create critical conditions (for example, local power surplus), and thus require the adoption of proper countermeasures.
As a consequence, significant changes in DS operation are expected. 
Active Distribution Networks (ADNs) must be conceived, where the need for smarter monitoring systems, supporting safe operation of the network becomes more and more evident.

Safe operation depends on the accurate knowledge of the electrical quantities in the grid \cite{14TPS_CelPeg_PilPisSul_DMSandDSSE}. 
Specific solutions, for the so called Distribution System State Estimation (DSSE), are required for DSs. 
Several DSSE techniques were presented in the literature (see, for instance, \cite{BarKel94IEEETransPowSysSE, LinTen96IEEEGTD_DistrFastDecoupled, WanSch04IEEETransPowSys, PauPegSul13TIM_EffBranchCurrentDSSE}).
It is worth noting that the DS monitoring process is made complex not only because of the scarcity of measurement devices but also due to 
the high number of nodes and the geographical extension of the networks in question.
In this context, several challenges are still open and new monitoring and management solutions are required.
They can be summarized as follows:

\begin{itemize}
\item DSSE methodologies able to efficiently treat the high number of nodes of the DSs;
\item measurement devices providing a time tag stating the reference time of the measurement results; 
\item communication infrastructures tackling the problem of efficiently collecting and 
coordinating the measurement results.
\end{itemize}

As for the methodologies, all those recently proposed in the literature are designed to efficiently face the estimation process so that the computational time can be limited. 
Among them, in particular Multi Area SE (MASE) methods are intended to address the issue of the 
high number of nodes (see, for instance, \cite{15-TIM_MASE-MusPauPegSulPonMon}).

As for the instrumentation, the Phasor Measurement Units (PMUs) are assuming a key role in protection and control systems. 
They measure voltage and current synchronized phasors (synchrophasors), along with frequency and rate of change of frequency (ROCOF), 
at a high reporting rate and with really high accuracy with respect to traditional supervisory control and data acquisition (SCADA) systems.
%Introduced in the late 1980s, PMUs have undergone a long standardization process.
The IEEE Standard for PMUs was released in 2011 by means of two documents \cite{IEEE_Std_C37.118.1-2011} and 
\cite{IEEE_Std_C37.118.2-2011}. An amendment, released in 2014, \cite{IEEE_Std_C37.118.1a-2014}, modifies or suspends some of the 
performance requirements specified in  \cite{IEEE_Std_C37.118.1-2011}.
These Standards aim to help in measuring dynamic signals that can occur in power systems.
Several algorithms have been proposed in the literature to address challenges due to the difficulty of measuring such signals. 
In particular, it is possible to mention \cite{14_tim_PM}, presenting an algorithm that allows the 
requirements of both the performance classes introduced in \cite{IEEE_Std_C37.118.2-2011} to be 
met. % simultaneously.

New ADN operators' requirements call for the modernization of the ICT infrastructure or the creation of a new one.
It is now necessary to collect and to process high volumes of data arriving from remote PMUs.
Cloud-based solutions can solve the non-trivial tasks related to storage, real-time computation and optimization of a
large amount of data as that generated in complex distribution systems.
In this scenario, the Cloud ensures a reliable environment with massive computational and storage capabilities. 
Moreover, it can elastically react to critical scenarios in which the need of resources for state estimation increases at a fast pace when specific events take place, thus ensuring an appropriate amount of resources are dynamically set only in critical time intervals, which is important from an economic point of view. 
Last but not least, the use of the IoT-related concept of virtualization on top of a cloud infrastructure boosts reusability of gathered data and is both evolution- and future-proof, since sensed data are decoupled from the details of the physical objects thanks to the virtualization layer.

The aforementioned challenges are open and intense research activities are expected.
Recently, communication infrastructures for the DSSE of active networks have been proposed, see for example \cite{15_SG_CloudRomano}.
However, at the distribution level, the variations of the signals during the operation can be impressively rapid and significant. 
For this reason, an infrastructure able to support flexible and scalable information access is a critical challenge. 
In this context, the paper presents a novel adaptive DSSE architecture based on virtualized PMUs   
and on Cloud-based IoT platform. 
The architecture is able to adapt the rate of the DSSE according to the decisions taken by properly 
distributed monitoring points. 
Several metrics can be applied to detect possible dynamics in the locally measured signals, but the underlying concepts
appear to be attractive for a continuous monitoring of the network status that is also 
bandwidth-saving and computationally efficient.
The validation of the method is performed on a small test system.
In the following, the overall procedure is described and the results obtained on a 13-bus DS are presented and discussed. 

\section{Background}

\subsection{Distribution System state Estimation}
DSSE is in charge of estimating the state of the network, in terms of node voltages or branch 
currents, starting from a few heterogeneous instruments, 
which measure several electrical quantities with different accuracies and reporting rates. 
The aim is to obtain a reliable picture of the network status so that the grid can be safely operated, 
because the accuracy of the measurement results is decisive for downstream decisions.
However, due to the lack of a sufficient number of measurement devices on the field, knowledge
obtained from a priori information has to be added to the measurements to make the system observable. 
This prior information is commonly referred to as pseudo-measurements in power systems literature.
As aforementioned, several methods for the DSSE have been presented in the literature, mostly based on a weighted least squares (WLSs) formulation.
In particular, recently, a new branch current DSSE (BC-DSSE) was proposed \cite{PauPegSul13TIM_EffBranchCurrentDSSE}, 
proving the same accuracy, but faster execution, as node voltage DSSE (NV-DSSE).
To  achieve  an  accurate  knowledge  of  the  network state, attention must be paid to the proper modeling of all the estimation problem elements. 
In addition, a detailed description of the measurement model is needed \cite{AsprouKyrAlbu14TIM_EffectWeights} and possible correlation existing in the measurements should be duly considered \cite{MusPauPegSul14TIM_EffCorrDSSE}.

Fast procedures can be able to properly use the high rate of PMUs measurements.
The full rate permits to have an up-to-date DSSE describing the dynamic of the system with the maximum resolution. 
This poses two main challenges. 
First of all, great computation flexibility is necessary, since estimations could pass from one 
every few seconds, as in SCADA systems, to one every $20\,\textrm{ms}$. 
Secondly, sending $50$, or more, measurement messages of at least $70$ bytes (only 
considering payload in 
simple PMU
configurations) from a single device could easily 
become an issue both for the communication and for the data storage, when economic and logistic
constraints are present.
 
Nevertheless, in normal operating conditions, DSs usually present near-steady-state signals. For 
this reason an adaptive estimation process can address the goal of efficiently monitor the network 
only in case of events.

\subsection{Phasor Measurement Units}
The PMUs were introduced in the late 1980s. 
%The first IEEE standard, the IEEE 1344, was published in 1995.
%In 2005, IEEE 1344 was replaced by IEEE C37.118-2005, dealing with issues concerning use of PMUs in electric power systems but lacking of considering power system dynamic activity. 
%A lot of PMUs on the field are compliant with this Standard. 
The last version of the Standard was released in December 2011, presenting two parts: the C37.118-1 \cite{IEEE_Std_C37.118.1-2011}, dealing with the phasor estimation processes, and the C37.118-2 \cite{IEEE_Std_C37.118.2-2011} dealing with the communications protocol.  
The Standard \cite{IEEE_Std_C37.118.1-2011} defines two performance classes, P and M, for 
protection and monitoring-oriented applications, respectively.
The  P-class  is  intended  for  applications requiring fast  measurement response  time,  especially  protections  in  power  system,  
while  the  M-class  should  be  considered  for  the  case  where  measurement  accuracy  is  of  crucial importance.  
A  standard  compliant  PMU  should  meet  all  the requirements  at  least  for  one  class.  
The  main  differences between  the  two  performance classes  are  the  test  conditions for steady state performance and the requirements for dynamic performance,  
especially   as  far   as  frequency  and  ROCOF estimation are concerned.
%The amendment \cite{IEEE_Std_C37.118.1a-2014} was released in March 2014 to either modify or suspend some of the performance requirements in \cite{IEEE_Std_C37.118.1-2011}.

The Standard \cite{IEEE_Std_C37.118.2-2011} introduces a method for real-time exchange of synchronized phasor measurement data between power system equipment.
A messaging protocol that can be used with any suitable communication protocol for real-time communication between PMU, phasor data concentrators (PDC), and other applications is introduced.
All the data provided by PMUs must be aligned by the PDC by means of the time-tag included in the measures, to create a correlation between measures made in the same time but from different measurement points of the network. 
The collected data can be stored for future analysis or used for real time application. In the latter case, 
the performance in terms of the latency and the bandwidth of the communication network are not negligible.
The Standard \cite{IEEE_Std_C37.118.2-2011} shows an example of the hierarchy in the synchrophasor network where the PMUs are connected to an local PDC, 
generally near or in the substation and the data are aggregated and sent to a next level of the data collection network as a corporate PDC or super PDC \cite{10_Naspinet}
to aggregate data across the utilities.

Required PMU reporting rates are: $10$, $25$, $50$ frames/s for $50\,\textrm{Hz}$ systems.
The actual rate shall be user selectable. 
Support for other reporting modes is permissible, and higher rates such as $100$ or 120 
frames/s and rates lower than 10 frames/s (such as 1 frame/s) are encouraged by the Standard. 
A proper use of such different reporting rates can significantly improve the efficiency of the 
monitoring process.

\subsection{Cloud IoT systems}
The IoT paradigm has been evolving towards the creation of a cyber-physical world where everything can be found, activated, probed, interconnected, and updated, so that any possible interaction, both virtual and/or physical, can take place. 

\begin{figure}[h!]
\centering
\includegraphics [width=0.8\columnwidth] {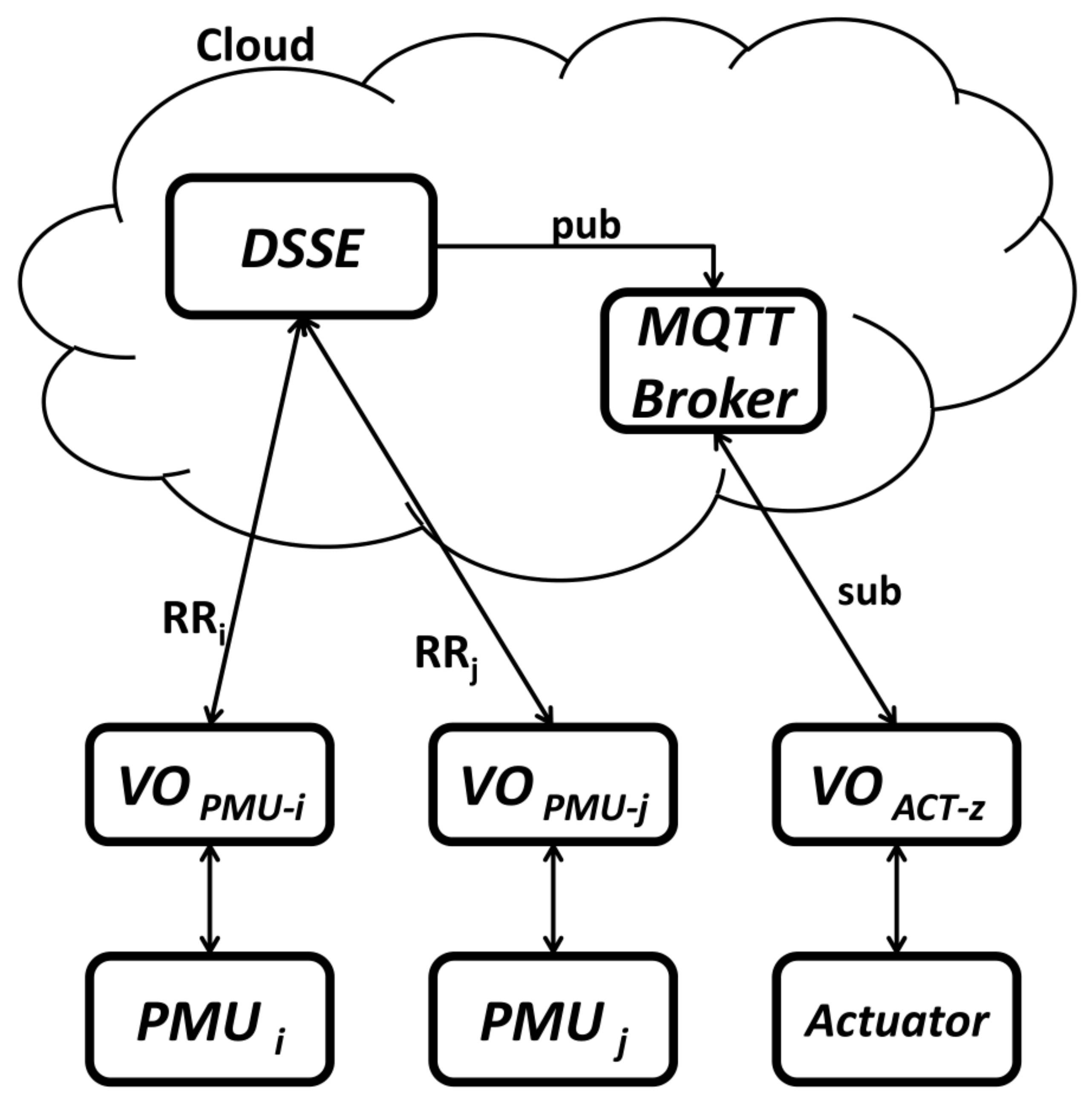}
\caption{Monitoring System Overview}
\label{Monitoring System}
\end{figure}

A crucial concept in this paradigm is the virtual object (VO), which is the digital 
counterpart of any real entity in the IoT. It has now become a major component of the current IoT 
platforms, supporting the discovery and mash up of services, fostering the creation of complex 
applications, improving the objects energy management efficiency, as well as addressing 
heterogeneity and scalability issues. Indeed, virtualization has the ability to: make 
heterogeneous objects interoperable through the use of semantic descriptions; enable them to 
acquire, analyse and interpret information about their context in order to take relevant decisions 
and act upon the virtual objects. Moreover, it enhances existing functionalities in the IoT 
promoting the creation of new addressing 
schemes, improving the objects mobility management efficiency, as well as addressing accounting 
and authentication issues. 

In addition, IoT platforms are more and more deployed in the cloud, as this approach allows for 
improving reliability, mash up of services, always on availability, elastic processing and memory 
resource provisioning. These features combined with the previously mentioned ones, make the 
virtualization and cloud computing the vital technologies for the future IoT solutions. 
In this paper, an IoT platform, namely Lysis (http://developers.lysis-iot.com/), has been implemented following the mentioned principles \cite{Girau2013}.

\section{Adaptive Distribution System State Estimation Architecture}

This paper presents an adaptive DSSE that is IoT based (the architecture scheme is in 
Fig.\ref{Monitoring System}).
Close to the physical measurement devices, proper Virtual Objects (VOs) have been designed.
A VO is defined as an entity that virtualizes device capabilities, so that any application can access or request its resources and functionalities in a reusable way, 
without knowing about the means (communications protocols and hardware primitives) that are needed to physically reach and retrieve information from the physical object.

In \cite{IEEE_Std_C37.118.2-2011} the commands, for the data collector, to control the streams of data provided by PMUs are defined. However, the functionalities are limited to enable and disable the real time stream. 
In this paper, the possibility to change the rate of the reporting of the measurements, between VO and the Cloud, is designed in order to save the bandwidth of the network while using the high reporting rate provided by the PMU only when is necessary. 

A distributed logic has been applied so that the DS monitoring can be adapted to possible events in the operating conditions.
The monitoring reporting rate must increase when a dynamic in the DS is detected, whereas the rate must decrease when the event has lapsed.

\subsection{Virtualized PMUs}
The considered measurement system is constituted PMUs (real PMU prototypes are exploited).
PMU measurements are sent with a GPS-synchronized timestamp to corresponding VOs, which are virtual containers implemented at the edge of the communication network. 
In our case, physical objects create a socket with the VO and send measured data according to \cite{IEEE_Std_C37.118.2-2011}. 
VOs are able to communicate with the Cloud at a given Reporting Rate ($RR$) and report only necessary information using REST APIs and the JSON format, thus abstracting from the PMU standard. 
Communications can either take place in $GET$ or $PUSH$ mode, which means that data can either be asked to the VO with a HTTP GET query or be sent automatically to a given location through HTTP POSTs. 
In the latter case, an appropriate trigger is set in the VO. 
In this paper, on the basis of the specific characteristics of the DSSE application considered, the focus will be on the case of an automatic HTTP POST.

As to the creation and deployment of the VOs, Lysis platform approach, hosted in the Cloud, is followed \cite{Girau2013}. 
It contains various VO templates each corresponding to a specific physical device. 
In our specific case, a number of templates are present corresponding to different physical PMUs (e.g. PMUs using different protocols such as IEEE 1344-95, IEEE C.37-118-2005, IEEE C.37-118.2-2011), since a different abstraction layer with the physical object is needed.  
Once the right interface is selected and the address of the VO location is given, the template is deployed in order to enable communication of the PMU with the applications implemented in the Cloud. This procedure, ensures the correctness of the VO setup and automates the procedure linked to the installation of new PMUs in the distribution network. It is important to highlight that the VOs are processes running at the communication network edge, so as to be close to the physical device. 

\subsection{Local process points}
Each PMU sends its measurements to the corresponding VO, which is a sort of gateway with context-awareness capabilities working as a protocol adapter and receiving data every $20\textrm{ms}$, that is at the maximum reporting rate suggested by \cite{IEEE_Std_C37.118.1-2011}. 
Thanks to context-awareness capabilities, VOs can perform proper processing in order to extract information on the state of the part of DS which is directly measured. 
In particular, the VO is able to adapt its output reporting rate towards the application, 
following a fixed policy for all the VOs in the network (but easily reconfigurable also on a 
per-VO basis). 

For the aim of this paper (as an example of detection methods of immediate understanding) the 
VO rate can be changed according to the monitoring of the 
following metrics with respect to given thresholds:
\begin{itemize}
\item the variation of RMS voltages between two consecutive input measurements (PMU measurements), 
 $\alpha$;
\item the variation of RMS voltages between two consecutive output measurements (VO measurements), 
 $\beta$;
\item the ROCOF value for each input measurement, $\gamma$;
\end{itemize}

Each of them can be used to command both the increase of the estimation rate and the decrease, according to the desired monitoring process.

\subsection{Adaptive DSSE rate}
The DSSE application is performed in the cloud at the application level {\footnote{With application level, we refer to the highest level of an IoT architecture as described in \cite{Atzori2010}} using the measurements received, at 
varying reporting rate, from the different VOs. 
By default, the output rate of the VO (and thus of the DSSE) follows a low reporting rate.
The measurements (with the corresponding timestamps), originated by the PMUs can, depending on the 
locally detected events, reach the DSSE at different rates (i.e. $50$, $25$, $10$ and $1$ 
frames/s). 
For this reason, the DSSE function is performed coordinating them at the highest reporting rate among the different measurement flows. As a consequence, a higher VO rate triggers a higher DSSE update rate, thus allowing 
to follow more accurately a faster event, even at nodes that are not directly monitored.

In this paper, the DSSE is performed using the fast branch-current state technique presented in \cite{PauPegSul13TIM_EffBranchCurrentDSSE}, exploiting 
the linearization of power injection pseudo-measurements to obtain a constant Gain matrix in WLS computation.
Nevertheless, it is worth noting that the adaptive architecture can be used for different 
estimation methodologies, without significant differences.

\section{Tests and Results}

\subsection{Test System}
The electric system used in the tests is composed by a sample of a DS derived from the IEEE 13-bus (Fig.\ref{fig:Test_System}).
The IEEE 13 bus radial distribution test feeder \cite{TestSys-IEEE} was proposed as a benchmark for the analysis of harmonic propagation in unbalanced networks.
For the purposes of this study, the topology and the loads of this network were considered as a starting point to design a simplified test network suitable for the proposed architecture.
In particular, the grid used for the test is, for the sake of simplicity, a totally balanced 
version of the IEEE 13 bus. 
In addition, a distributed generator was collocated on the network, at the node 34, so that 
presence of DER can be taken into account.

\begin{figure}[!h]
\centering
\includegraphics[scale=0.18, clip=true, trim=10 0 0 10]{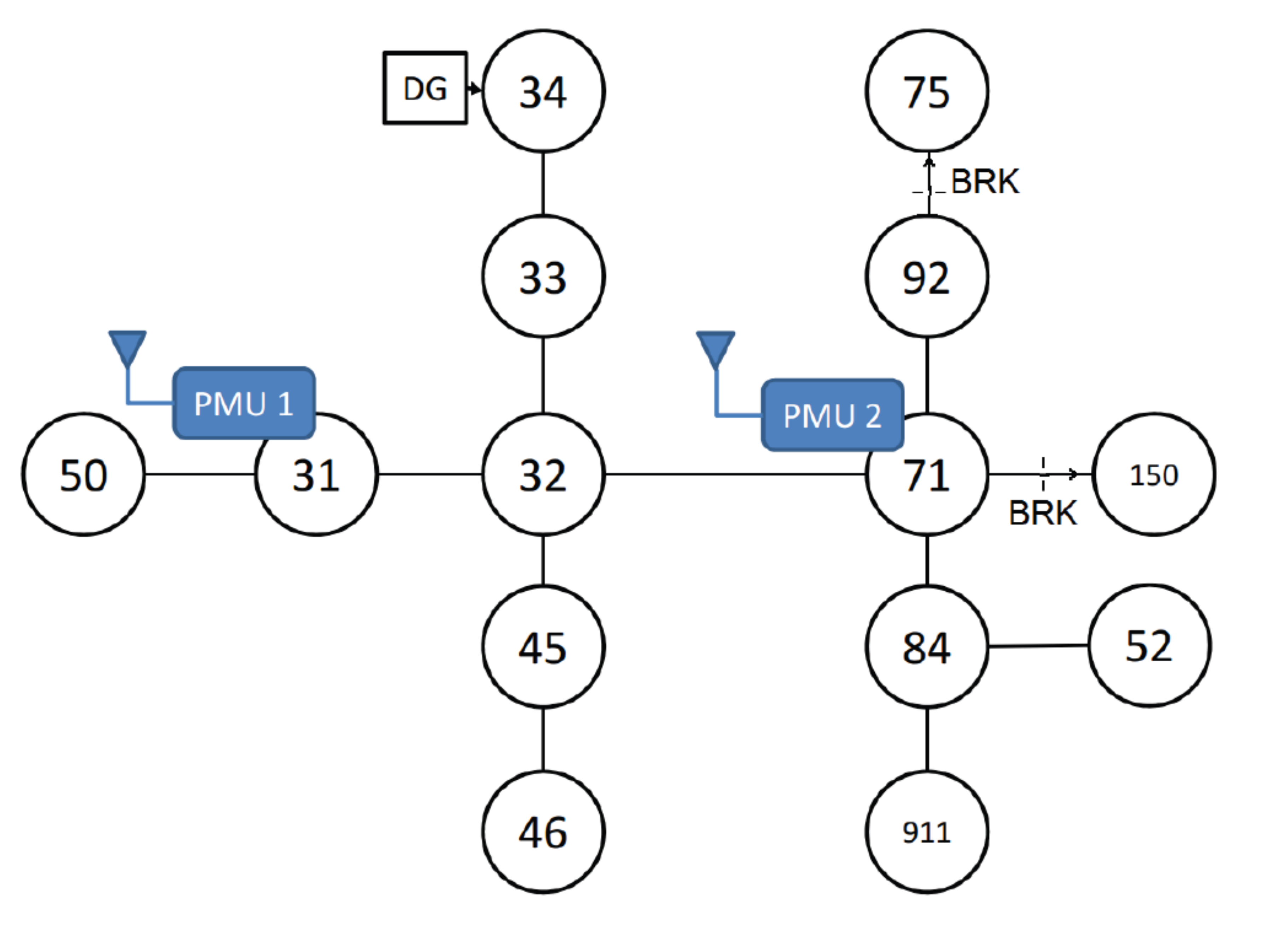}
%\vspace{-0.2cm}
\caption{Test system}
\label{fig:Test_System}
%\vspace{-0.3cm}
\end{figure}

The implementation of the network was carried out with the PSCAD/EMTDC software \cite{Pscad}, a well known 
design and simulation tool to model power systems, acting as a graphical user interface to the 
EMTDC simulation engine. The tool allowed the simulation of different events and dynamics, at 
different locations.

\subsection{Monitoring System}

For a realistic measurement scenario, two PMUs were placed in two points of common coupling of the network. 
In particular, phasor voltages at nodes 31 and 71 are measured and the synchrophasors are computed, along with frequencies and ROCOFs.

Each PMU prototype is implemented using the real-time embedded controller NI-cRIO (Fig.\ref{fig:Prototype}). 
The system is a reconfigurable device and, 
in the adopted configuration, is composed by a real-time controller NI-9014, a Field Programmable Gate Array (FPGA) module embedded in a chassis NI-9113, a NI 9215 16-Bit Simultaneous Analog Input Module, and a time synchronization module NI-9467. The time synchronization module is a GPS receiver that offers an accurate 
time source (accuracy $\pm$ 100 ns) to synchronize the embedded clock of the FPGA module and to provide the Coordinated Universal Time (UTC) 
to the real time controller. For this reason, every sample acquired by the PMU can be tagged directly at FPGA level with the UTC timestamp 
while the real-time controller is in charge of higher level phasor, frequency and ROCOF computations, and of the data frame encapsulation 
and transmission.

\begin{figure}[!h]
\centering
\includegraphics[scale=0.9, clip=true, trim=40 50 40 50]{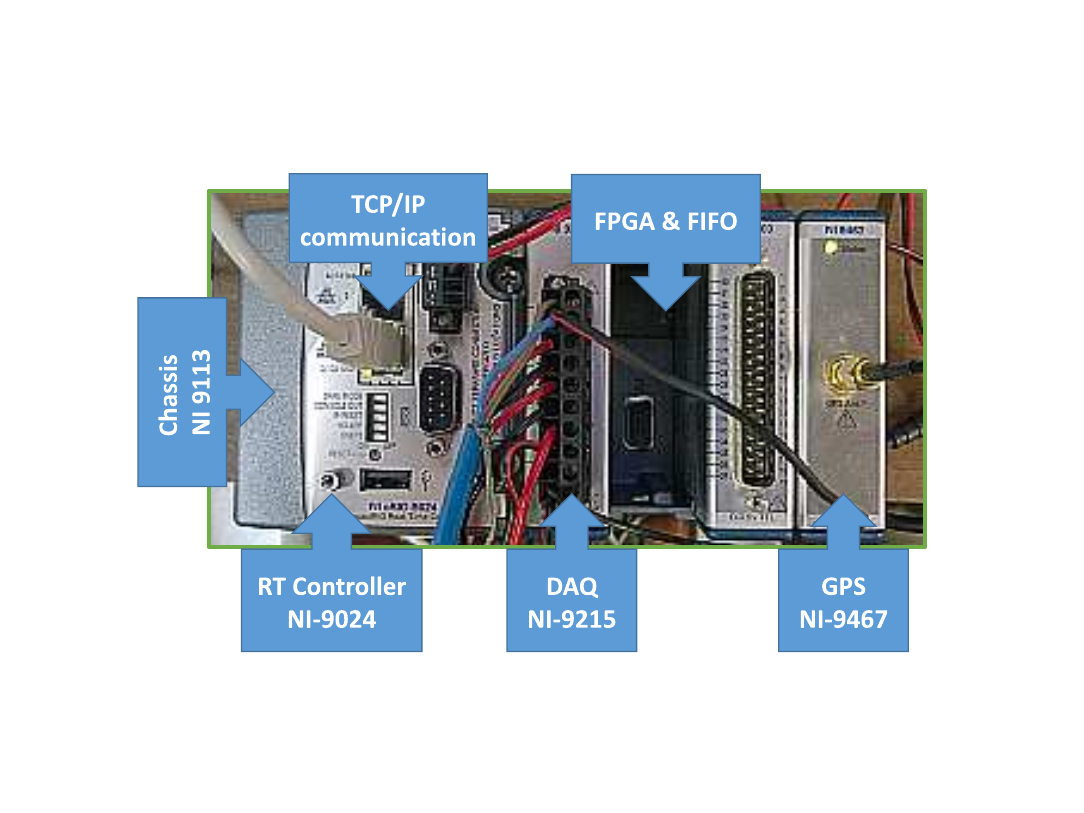}
%\vspace{-0.2cm}
\caption{PMU Prototype}
\label{fig:Prototype}
%\vspace{-0.3cm}
\end{figure}

The PMUs are configured to run the fast P-class estimation algorithm proposed in \cite{15TIM_DistrPMU} at 50 frames/s.
P-class algorithm has been chosen because it is faster to react to dynamic conditions and because its accuracy and latency characteristics do not change with reporting rate.

For the aim of this paper, two PMU prototypes, synchronized by means of GPS, are modified to 
compute synchronized measurements on the pre-stored test signals obtained by PSCAD simulations and 
corrupted by a level of noise of $70$ dB, corresponding to a possible noise level of the data acquisition stage. 

The computed synchrophasors are sent in real-time following the IEEE C37.118.2 message format.

As example of decision-making process at each input measurement message, the VO checks:

\begin{itemize}
\item if $\alpha > 2 \%$; 
\item if $\beta > 2 \%$ 
\item if $\gamma > 5$ Hz/s 
\end{itemize}
to decide whether to increase the output rate to the maximum rate, i.e. $RR=50$ frames/s.
Furthermore, the VO monitors if $\beta$ drops below $0.1\,\%$ to slow down the $RR$ 
progressively, and with a given inertia, to $25, 10, 1$ frames/s.
A PMU, in general, is characterized by its measurement accuracies in presence of steady-state and 
dynamic conditions and this is the reason why different thresholds have been used for $RR$ 
variations, keeping into account the device accuracies. 

In the considered scenario, the state estimation, whose speed is changed when certain events are 
triggered on a node, can be compared with the DSSE outputs at a fixed rate, that is the most 
simple 
case, relying on simple and constant PMU configuration.

\subsection{Results}
In order to test the dynamics of the measurement system, several events have been generated and 
measured on the network.

In particular, for the sake of simplicity, the results concerning some possible operations on two 
loads are reported. 
The events are triggered in given moments of the operation of the network: 
\begin{itemize}
\item breaker at node 75 opens at $13.5\,\textrm{s}$ and closes at $15.5\,\textrm{s}$;
\item breaker at node 150 opens at $8.2\,\textrm{s}$ and closes at $17.6\,\textrm{s}$.
\end{itemize}
 
Meanwhile PMUs are continuously monitoring, measurements are collected by the corresponding VOs 
and then VOs check the aforementioned metrics. Each VO is expected to detect variations that 
affects the monitored voltage.
%For the aim of this paper, the variations of the rms values of the measured voltage are monitored. 
%In case such variations exceed a certain threshold then a thicker monitoring is prompted.
%Both variations between close measured values are monitored and the trend of variations among several measurements.  

\begin{figure}[!h]
\centering
\includegraphics[scale=0.47, clip=true, trim=10 0 0 20]{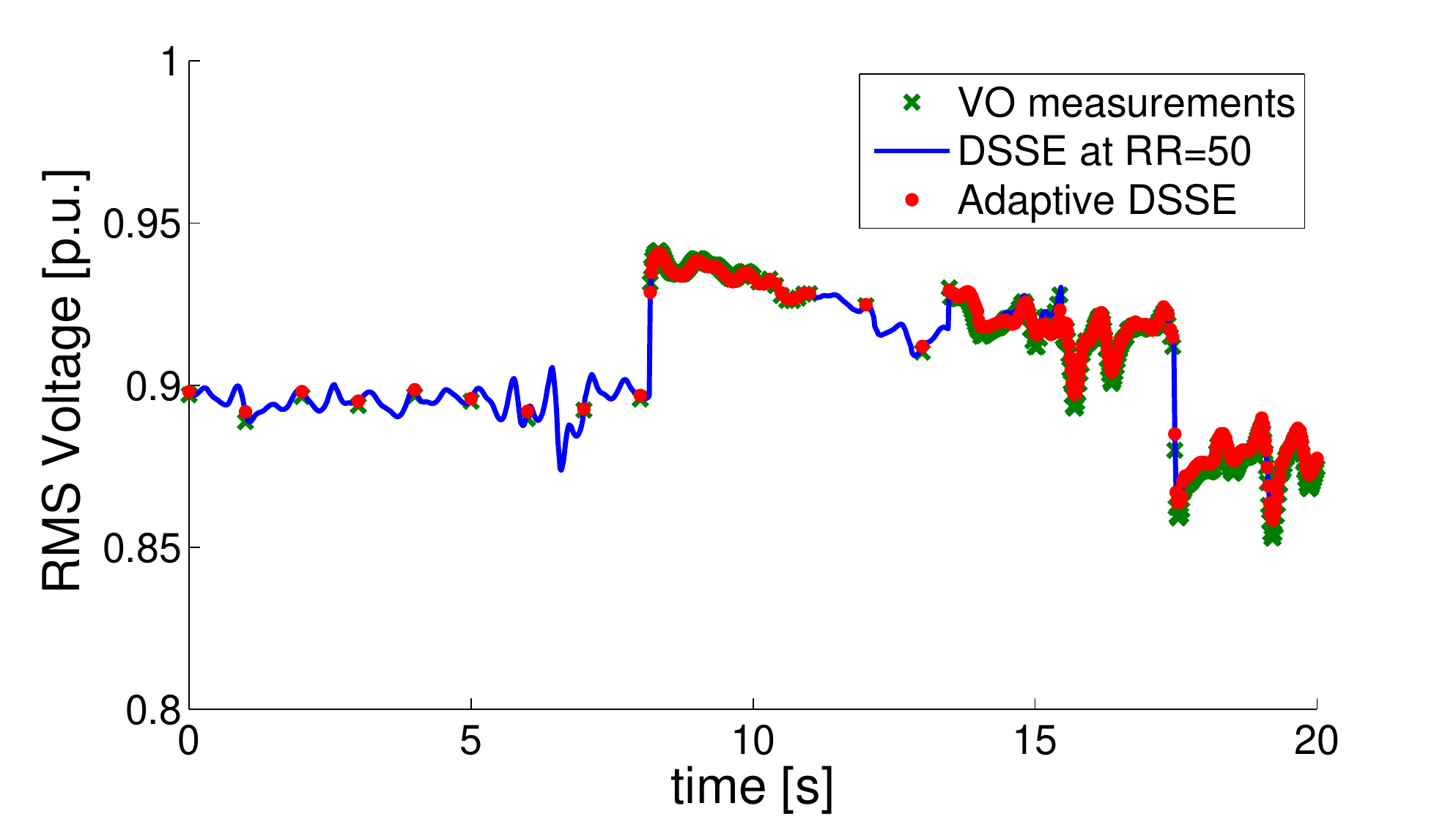}
\vspace{-0.2cm}
\caption{RMS voltage estimation for node 71 evaluated with Adaptive DSSE.}
\label{fig:Nodo5}
\vspace{-0.3cm}
\end{figure}

Fig. \ref{fig:Nodo5} shows, for instance, the results obtained at node 71. 
The adaptive DSSE results are indicated by red dots. 
The measurements sent by the VO are represented by green ``x`` and the DSSE results at the 
constant reporting rate 
$RR=50$ frames/s are also reported to serve as comparison.
For the sake of clarity, a continuous blue line is used to connect the thin grid points obtained on the 20 ms resolution scale. 

The actual reporting rate clearly follows the dynamic of the RMS voltage and the VO correctly detects the 
fast transient (a breaker opened), thus triggering the DSSE to take frequent snapshots of the 
network status, only when a interesting phenomenon is under investigation.
Furthermore, the dynamics due to the other breaker operating on the other load and the reclosing 
of the breaker are correctly identified.

\begin{figure}[!h]
\centering
\includegraphics[scale=0.47, clip=true, trim=10 0 0 20]{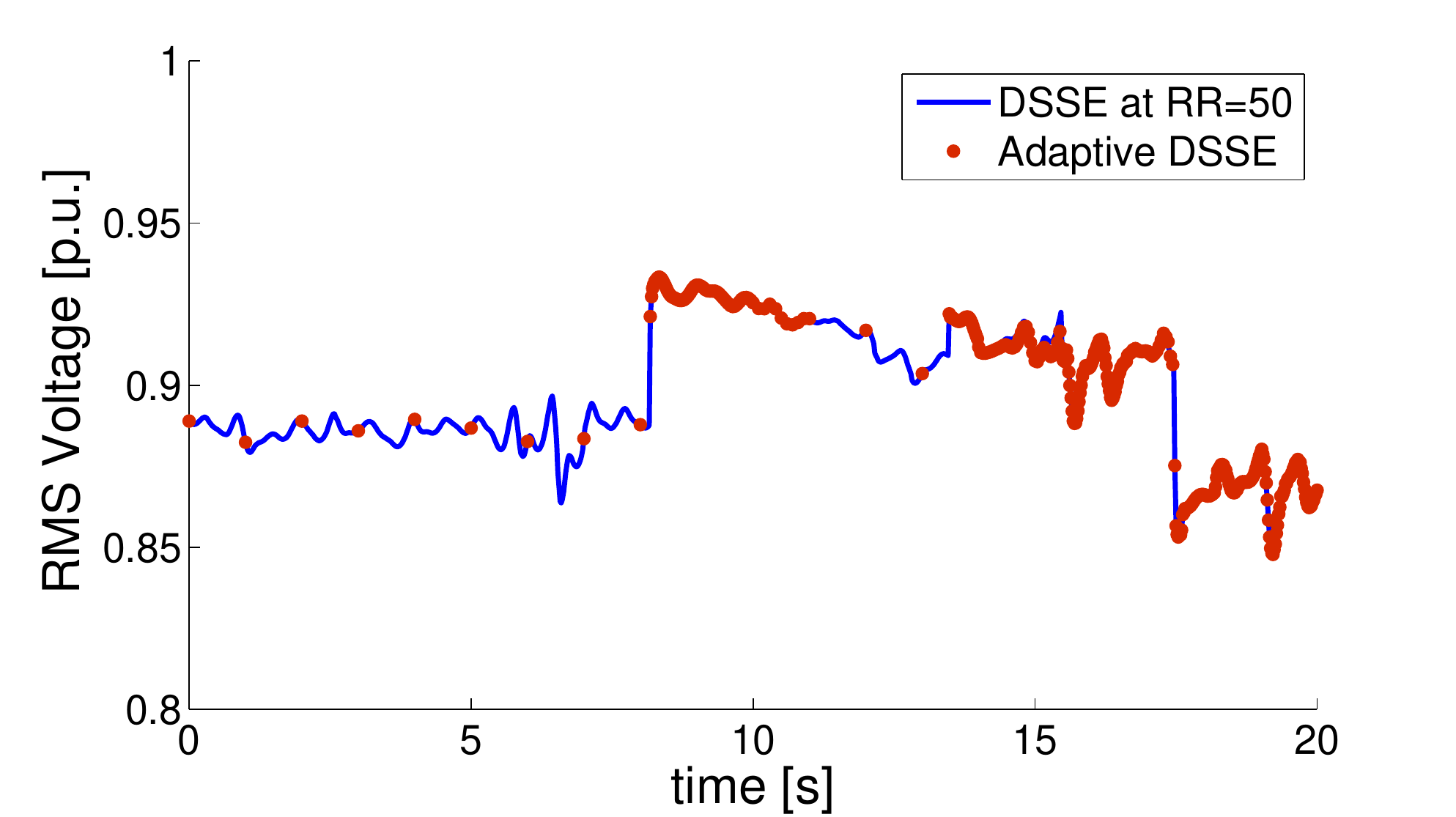}
\vspace{-0.2cm}
\caption{RMS voltage estimation for node 150 evaluated with Adaptive DSSE.}
\label{fig:Nodo7}
\vspace{-0.3cm}
\end{figure}

Fig. \ref{fig:Nodo7} reports the results of the DSSE in terms of the RMS voltage estimation at 
node 150 (that is not directly monitored) when both the maximum $RR$ given by PMUs and the 
Adaptive DSSE are used. 
The disconnection of the load leads to an increase of node voltage amplitude: the 
DSSE promptly reacts to such event and is able to follow the dynamic of the unmonitored node thank to the adaptive policy of VOs and to the variable DSSE reporting rate.

\section{Conclusions}
This paper presents an auto adaptive distributed system state estimation in an IoT cloud-based system. 
The methodology is adaptive in the rate of execution of the estimation process.
The rate varies depending on the indications of the PMU measurement system.
The major features are: the virtualization of the PMUs with the virtual objects running on the edge of the communications network and the implementation of a logic at the virtual object so that the transmission rate of the measurements and the estimation process can be adapted on the basis of given metrics. 
The performance analysis were conducted in an example distribution network derived from the IEEE 
13-bus.
The obtained results prove the goodness of the methodology with accurate tracking of the events 
occurring in the network along with an efficient management of the transmission rate compared with 
the case of a full rate monitoring architecture.

\section*{Acknowledgement}

This work has been supported by Regione Autonoma della Sardegna, L.R. 7/2007: Promozione della 
ricerca scientifica e dell'innovazione tecnologica in Sardegna, annualit\`{a} 2012, CRP-60511".

\bibliographystyle{IEEEtran}
\bibliography{./bibliography/IEEEabrv,./bibliography/stateestimationandpowerflow,./bibliography/synchrophasors,./bibliography/meterplacement,./bibliography/references} 

\end{document}